\documentclass[12pt]{iopart}
\usepackage{iopams}
\usepackage{graphicx,epsf}
\usepackage{bm}

\usepackage{hyperref}

\begin{document}

\jl{14}
\title[2D continuous spectrum of SAW in the presence of a magnetic island]{2D continuous spectrum of shear Alfv\'en waves in the presence of a magnetic island}

\author{Alessandro Biancalani}
\address{Max-Planck-Institut f\"ur Plasmaphysik, Euratom Association, D-85748 Garching, Germany\\
in collaboration with Max-Planck-Institut f\"ur Sonnensystemforschung, Katlenburg-Lindau, Germany.}
\author{Liu Chen}
\address{Institute for Fusion Theory and Simulation, Zhejiang University, Hangzhou, People's Republic of China\\
Department of Physics and Astronomy, University of California, Irvine, CA 92697-4575, USA}
\author{Francesco Pegoraro}
\address{Department of Physics, University of Pisa, 56127 Pisa, Italy}
\author{Fulvio Zonca}
\address{Associazione Euratom-ENEA sulla Fusione, C.R. Frascati, C.P. 65 - 00044 Frascati, Italy}
\date{\today}
\begin{abstract}

The radial structure of the continuous spectrum of shear Alfv\'en modes is calculated in the presence of a magnetic island in tokamak plasmas.
Modes with the same helicity of the magnetic island are considered in a slab model approximation. In this framework, with an appropriate rotation of the coordinates the problem reduces to 2 dimensions. Geometrical effects due to the shape of the flux surface's cross section are retained to all orders. On the other hand, we keep only curvature effects responsible of the beta-induced gap in the low-frequency part of the continuous spectrum. New continuum accumulation points  are found at the O-point of the magnetic island. The beta-induced Alfv\'en Eigenmodes (BAE) continuum accumulation point is found to be positioned at the separatrix flux surface. The most remarkable result is the nonlinear modification of the BAE continuum accumulation point frequency.

\end{abstract}
\pacs{52.55.Tn, 52.35.Bj}
\submitted
\maketitle

\section{Introduction}
\label{sec:start}

Plasma stability is one of the crucial issues for fusion devices. Shear Alfv\'en instabilities can resonate with energetic particles and are therefore particularly dangerous for a burning plasma~\cite{Rosenbluth75,Mikhailovskii75,Chen89,Fu89,Zonca06b,chen07}. Shear Alfv\'en waves (SAW) are electromagnetic plasma waves propagating with the characteristic Alfv\'en velocity $v_A=B/\sqrt{4\pi\varrho}$ ($B$ is the magnetic field and $\varrho$ the mass density of the plasma). One of the main damping mechanisms of shear-Alfv\'en modes in nonuniform plasmas is continuum damping~\cite{Grad69,hasegawa74,chen74}, due to singular structures that are formed at the SAW resonant surfaces. Due to nonuniformities along the field lines in toroidal geometry, gaps appear in the SAW continuous spectrum~\cite{kieras82}. The mechanism is similar to that which creates forbidden energy bands for an electron traveling in a periodic lattice~\cite{cheng85,Zonca92,Chen95}. Two types of collective shear Alfv\'en instabilities exist in tokamak plasmas~\cite{Zonca06a}: Energetic Particle continuum Modes (EPM)~\cite{chen94}, with frequency determined by fast particle characteristic motions, and discrete Alfv\'en Eigenmodes (AE), with frequency inside SAW continuum gaps~\cite{cheng85}. Discrete AE are practically unaffected by continuum damping~\cite{Zonca06b,chen07}. For this reason, the importance of understanding the continuous spectrum structure is clear, if one faces the tokamak stability problem and its potential impact on reaching the ignition condition.

SAW nonlinear dynamics is a research topic with many open issues.
In the special case of uniform plasmas a peculiar state exists, called the \emph{Alfv\'enic state}, which arises whenever we can assume ideal MHD, plasma incompressibility and the validity of the Wal\'en relation ($\delta v /v_A = \pm \delta B / B$, where $\delta v$ and $\delta B$ are the perturbed velocity and field)~\cite{Alfven50,Elsasser56,Hasegawa89}. In such a case, the nonlinear effects due to Maxwell and Reynolds stresses cancel and give a self-consistent nonlinear state. On the other hand, in nonuniform tokamak plasmas, with non-ideal effects such as resistivity or finite plasma compressibility, the Alfv\'enic state conditions are violated and SAW are characterized by a rich nonlinear dynamics.
The study of the nonlinear modification of the SAW continuous spectrum due to MHD fluctuations can be of practical interest in realistic tokamak scenarios. In fact, in a tokamak plasma, the SAW continuous spectrum is modified by the interaction with low-frequency MHD fluctuations, such as magnetic islands, which are formed when the original sheared equilibrium magnetic field lines break due to non ideal effects (in particular finite resistivity) and reconnect with different magnetic topology~\cite{furth63}. Since the typical island frequency and growth rate are much lower than the SAW oscillation frequency, we can model the equilibrium magnetic field as the sum of a tokamak axisymmetric part plus a quasi-static helical distortion due to the magnetic island.

Here, we derive the fluid theoretical description of the SAW continuum structure in the presence of a finite size magnetic island~\cite{BiancalaniEPS08,Tuccillo09} in finite-$\beta$ plasmas, keeping into account only toroidal effects due to geodesic curvature, which are responsible for the beta-induced Alfv\'en Eigenmodes (BAE) gap in the low frequency part of the SAW continuous spectrum~\cite{chu92,turnbull93,zonca96}. We adopt a linear ideal MHD model, and we consider only shear Alfv\'en modes with the same helicity of the magnetic island. In this framework, with an appropriate rotation of the coordinates the problem reduces to 2 dimensions. Since we are interested in the structures of the SAW continuous spectrum, characterized by local ``radial'' singular behaviors, the problem can be further reduced to one dimension. The local differential equation describing the nonlinear SAW continuum structure is solved numerically with a shooting method code in the whole spatial range of interest (both outside and inside the island), and analytically near the O-point and at the island separatrix, where the equilibrium quantities can be approximated in simple form. Note that what we are calculating is essentially the distortion of the SAW continuous spectrum in a one-dimensional system (1D-slab or cylinder) due to the presence of a magnetic island. For this reason, for given island helicity, we refer to the structures derived here as ``2D continuous spectrum of SAW'' in the presence of a magnetic island. Note also that the problem investigated in this work is somewhat different with respect to that studied in~\cite{Biancalani2010prl}, where modes with different helicities from that of the magnetic island are considered\footnote{Ref.~\cite{Biancalani2010prl} analyzes this problem with a simple representation of the magnetic island as a straight flux-tube with non-circular cross-section.}, yielding 3D structures of the SAW continuous spectrum.

As suggested by the magnetic field line helicity behavior~\cite{swartz84,buratti05nufu}, in an equilibrium with a magnetic island the separatrix flux surface plays an important role, hosting the BAE Continuum Accumulation Point (BAE-CAP), which - without island - was positioned at the rational surface. The degeneracy in frequency of even and odd modes is removed by inhomogeneities along the field lines, causing a splitting between the continuous spectrum branches with different parities. As a consequence, the BAE-CAP is also split in different frequencies, depending on the mode number and parity of the eigenfunction. Several branches of the nonlinear SAW continuous spectrum stem from the BAE-CAP. Inside the island, they reach  continuum accumulation points at the magnetic island O-point (MiO-CAP), while outside the island and far from the separatrix, they asymptotically approach the typical spatial dependence of the SAW continuum in a sheared magnetic field in the absence of the island.

Understanding the modification of the SAW continuous spectrum due to the presence of a magnetic island has potential implications in explaining stability of Alfv\'en instabilities in tokamak plasmas.
Modes in the BAE frequency range have been observed in the Frascati Tokamak Upgrade (FTU)~\cite{buratti05nufu,BiancalaniIAEA08} in the presence of an $(m,n)$ = (-2,-1) magnetic island, where $m$ and $n$ are, respectively, the poloidal and toroidal mode number. A theoretical analysis showed that these modes can be interpreted as BAE modes, when thermal ion transit resonances and finite ion Larmor radius effects are accounted for, with good agreement of measured and calculated frequencies in the small magnetic island amplitude limit~\cite{annibaldi07}. In fact, measured frequencies were found to depend on the magnetic island amplitude as well~\cite{buratti05nufu}, consistently with the dependence of the BAE-CAP on the magnetic island size resulting from our theory. The modes were observed only when the magnetic island size was over a certain critical threshold~\cite{buratti05nufu}. Later on, similar observations have been reported in other tokamaks (see, for instance, the observations in HL-2A~\cite{Wei2010}). This work derives and discusses the structures of the SAW continuous spectrum in tokamak geometry and in the presence of a finite size magnetic island. As mentioned above, the nonlinear modification of the BAE dispersion relation due to the presence of a magnetic island is not treated here and will be presented in a subsequent paper. The scheme of the paper is the following. In Sec.~\ref{sec:theory} the equilibrium magnetic field and the model equations are given; meanwhile, a convenient a coordinate system is defined inside and outside the magnetic island. In Sec.~\ref{sec:inside} we solve the problem for the continuous spectrum both numerically, in the whole region inside the magnetic island and analytically, near the O-point, obtaining the value of the MiO-CAP. Similarly, in Sec.~\ref{sec:outside} we solve the problem for the region outside the magnetic island and compare the numerical solution with the asymptotic analytical solution far from the separatrix. The frequency of the BAE-CAP is also calculated as a function of the island size and mode numbers and parity. Sec.~\ref{sec:conclusions} is devoted to a summary of the novel obtained results and their possible application to the study of BAE in the presence of a magnetic island.

\section{Equilibrium and model equations}
\label{sec:theory}

\subsection{Coordinate system}

\begin{figure}[t!]
\begin{center}
\includegraphics[width=0.6\textwidth]{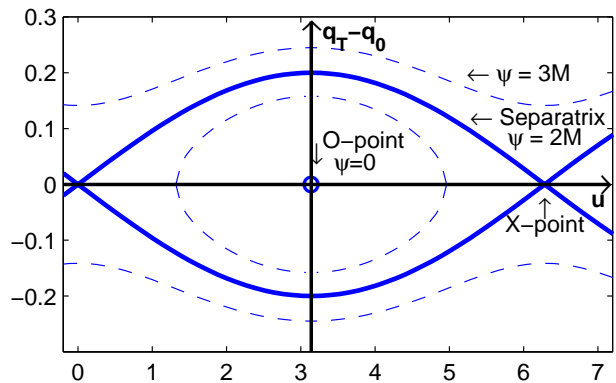}
\caption{Island coordinate system. The horizontal axis $q_T-q_0 = 0$ is the rational surface of the island. In this example, the amplitude of the island is chosen as $M =10^{-2}.$}\label{fig:island}
\end{center}
\end{figure}

We consider a tokamak geometry for a torus with major radius $R_0$. The equilibrium is made of an axisymmetric tokamak magnetic field with a component $B_{tor}$ in the toroidal direction $\zeta_T$ and a component $B_{pol}$ in the poloidal direction $\theta_T$, plus an helical perturbation in the radial direction $r_T$, generating a magnetic island. The magnetic island is located at a flux surface with minor radius $r_0$. Here, the subscripts $T$ denote tokamak coordinates. We consider the region in the proximity of the rational surface of the magnetic island $q_T = q_0 = m_{isl}/n_{isl}$, where $m_{isl}$ and $n_{isl}$ are, respectively, the poloidal and toroidal mode numbers of the magnetic island perturbation and $q_T= r_T B_{tor}/(R_0 B_{pol})$ is the safety factor. In this region, the toroidal magnetic field is assumed to be dominant and constant in space, $B_{tor} = B_0 /\gamma$, and the poloidal magnetic field is considered to vary linearly with $q_T$. Here $B_0$ is the axisymmetric magnetic field amplitude, $\gamma = \sqrt{1+ \varepsilon_0^2 / q_0^2}$, and $\varepsilon_0 = r_0 / R_0$. We adopt a slab model with coordinates ($q_T,u,\zeta$), where $u$ is defined in Eq.~\ref{eq:coordinates1}, by applying a rotation of the coordinates in the ($\theta_T,\zeta_T$) plane, as shown in Fig.~\ref{fig:island}. In this model, the coordinate $\zeta=(\zeta_T + \varepsilon_0^2 \, \theta_T /q_0)/(q_0 \gamma^2)$ is the coordinate of translational symmetry. 
The equilibrium magnetic field is ${\bm B}=B_{q_T,ph} \bm{\hat{r}}_T +B_{u,ph} \bm{\hat{u}}+B_{\zeta,ph} \bm{\hat{\zeta}}$, where the axisymmetric field components are $B_{u,ph} = B_0 \varepsilon_0 (q_T-q_0)/(q_0^2 \gamma^2)$ and $B_{\zeta,ph} = B_0$ (the \emph{physical} components are defined in detail in~\ref{appendix:metric}). The \emph{constant}-$\psi$ approximation is also adopted, assuming that the magnetic island field is $B_{q_T,ph} = B_{isl} \sin u$, with $B_{isl}$ constant.

The flux surfaces of this equilibrium are labelled by $\psi$, where the coordinates $\psi$ and $u$ are defined by:
\begin{equation}\label{eq:coordinates1}
\psi = (q_T-q_0)^2 /2 + M (\cos u +1) \; , \; \; u =n_{isl}(\zeta_T - q_0 \theta_T)
\end{equation}
The X-points of the magnetic island are at $(q_T-q_0,u)=(0,0)$ and $(0,2 \pi)$ and the O-point at $(0,\pi)$.  $M$ is a constant with value $M=(q_0 |s| /n_{isl}) (B_{isl}/ B_{pol,0})$, determined by the condition ${\bm\nabla}\psi \cdot \bm B =0$, where $s$ is the magnetic shear and $B_{pol,0}$ is the poloidal magnetic field evaluated at the rational surface.

In the slab model approximation, the plasma inside the magnetic island is a straight flux tube with length $Z_0 = \gamma q_0 R_0$. The magnetic axis of the flux tube, directed along $\hat{\bm{\zeta}}$, and the O-point of the magnetic island are at $\psi = 0$ and the separatrix is labeled by $\psi = \psi_{sx}= 2 M$. An appropriate set of cylinder-like coordinates $(\rho,\theta,\zeta)$ is defined here to describe the region inside the magnetic island (see~\ref{appendix:metric}), with:
\begin{equation}\label{eq:coordinates2}
\rho = \frac{r_0}{q_0 s} \sqrt{2 \psi} \; , \; \; \theta  =  \arccos (\sqrt{M(\cos u + 1)/\psi} \,)
\end{equation}
With these definitions, the magnetic axis is at $\rho=0$ and the separatrix radius is $\rho =\rho_{sx}$, which corresponds to the magnetic island  half-width, $W_{isl}$, given by the Rutherford formula~\cite{Rutherford73}:
\begin{displaymath}
\frac{W_{isl}}{r_0} = 2 \sqrt{\frac{B_{isl}}{q_0 s n_{isl} B_{pol,0}}} = \frac{\rho_{sx}}{r_0} = \frac{2}{q_0 \gamma n_{isl}} \sqrt{1-e}
\end{displaymath}
with $e$ defined below in this paragraph.
The angle $\theta$ is defined in the domain $(0,\pi/2)$ and extended to $(0,\pi)$ by reflection symmetry w.r.t. $\theta=\pi/2$. Further extension to $(0,2\pi)$ is obtained by reflection symmetry for $\rho \leftrightarrow -\rho$.
With this definition, $\theta$ has values $0,\pi$ at the rational surface $q = q_0$. We also point out that the flux surface's cross section in the $(\rho,\theta)$ plane and in the proximity of the O-point is an ellipse, with eccentricity $e = 1 - M n_{isl}^2 \gamma^2 /s^2$. Typical magnetic islands in tokamak experiments have values of eccentricity close to $e\simeq 1$. Outside the island, the flux surfaces have the same topology of the original axisymmetric equilibrium, and therefore the coordinate system ($\rho,u,\zeta$) is appropriate to describe the problem (see Eq.~\ref{eq:coordinates1}).

\subsection{Equilibrium magnetic field and model equations}

The equilibrium magnetic field in the coordinates ($\rho,u,\zeta$) is described by the contra-variant physical components: $B^\rho_{ph} = 0$, $B^u_{ph} =  2 \varepsilon_0 B_0 \sqrt{M} P / (q_0^2 \gamma^2)$, $ B^\zeta_{ph}  = B_0$. Here $P = \sqrt{L^2 + (1-e)(\sin^2 u)/4}$, where $L = \sqrt{x^2 - (\cos u + 1)/2}$ and $x = \rho/\rho_{sx}$. The contra-variant \emph{physical} components of a vector $\bm{V}$, in a basis $\{ \bm{g}_\rho ,\bm{g}_u, \bm{g}_\zeta \}$ are defined here as the contra-variant components rescaled with the length of the correspondent basis vector, e.g. $V_{ph}^\rho = V^\rho |\bm{g}_\rho|$ (see~\ref{appendix:metric} for further details). Similarly, in the coordinates defined inside the island we have $B^\rho_{ph} = 0$, $B^\theta_{ph} = 2 \varepsilon_0 B_0 \sqrt{M} \sqrt{a} \, x / (q_0^2 \gamma^2)$, $ B^\zeta_{ph}  = B_0$. The function $a$ is defined as $a = \sin^2\theta + \cos^2\theta (1-e)  F^2$, with $ F= \sqrt{1-x^2 \cos^2\theta}$.

The safety factor outside the island, $q_{out}$, is defined as the flux-surface average of the ratio of the toroidal magnetic field and the sum of the poloidal and radial magnetic fields. The flux-surface average is performed separately on both sides of the magnetic island, with $r > r_0$ and with $r < r_0$, along the coordinate $y_{out} = \int_0^{\theta_T} r_0 d\theta_T \sqrt{1+B_{q_T,ph}^2 / B_{\theta_T,ph}^2}$. By definition, we have:
\begin{equation}\label{eq:q-out}
q_{out} = \mathop{\int\mkern-20.8mu \circlearrowright}
\frac{d y_{out}}{2 \pi R_0} \frac{B_{tor}}{ B_{ph}^{y_{out}} } =
\frac{1}{2 \pi n_{isl}} \int_0^{2 \pi q_0 n_{isl}} \hspace{-1 cm} \frac{du}{( 1-2 \sqrt{M} P / q_0 )   }
\end{equation}
Far from the separatrix $(x>1)$, the safety factor reaches asymptotically the linear behavior typical of the slab model without island. When the magnetic island amplitude $M$ vanishes, this behavior is recovered for all $x$. In fact, in this case we have that $2 \sqrt{M} P \rightarrow (q_T - q_0)$ and therefore $q_{out}\rightarrow q_T$. An important result is that the safety factor at the separatrix $(x=1)$ is not $q_0$, but there is a small difference, proportional to the magnetic island size. This difference is positive on one side of the island, where $q_T > q_0$, and negative on the side where $q_T < q_0$. The absolute value of this difference can be calculated at the separatrix from Eq.~\ref{eq:q-out}, in the limit $M \ll 1$. We obtain:
\begin{equation}\label{eq:q-sx-approx}
 q_{out,sx} -q_0 = \frac{4 \sqrt{M}}{\pi} \simeq 1.27 \sqrt{M}
\end{equation}
This result has important implications for the modification of the value of the BAE continuum accumulation point, which is located at the separatrix. This island-induced frequency shift of the BAE-CAP can be calculated in an approximated form as: $\Delta\omega \simeq k_\| v_A = v_A m (q_{out,sx}-q_0)/(q_0 Z_0) \simeq  1.27 \sqrt{M} (v_A/Z_0) (m/q_0)$, where $m$ is the poloidal mode number. In fact, we will see in Section~\ref{subsec:separatrix} that the BAE continuum accumulation point has not the same frequency as without island, but a nonlinearly modified frequency which is proportional to $\sqrt{M}$, consistently with the nonlinearly modified value of the safety factor given by Eq.~\ref{eq:q-sx-approx}.

Similarly, the safety factor inside the magnetic island can be defined
as the average of $ Q = \rho B_0 / ( Z_0 B^\theta ) $ over $\theta$, with $B^\theta = B^\theta_{ph} F \sqrt{1-e} / (\sqrt{a} \,\rho)$ (see also reference~\cite{Biancalani2010prl}):
\begin{equation}\label{eq:safety-factor}
q_{in} =  \frac{2}{\pi} \frac{\gamma}{|s| \sqrt{1-e}} \mathtt{K}(x) =  \frac{2}{\pi} \frac{1}{\sqrt{M} n_{isl}} \mathtt{K}(x) \; ,
\end{equation}
where we have introduced the complete elliptic integral of the first kind $\mathtt{K}(x) = \int_0^{\pi/2} d\theta /F$. A similar definition for the safety factor was given in Ref.~\cite{swartz84}, but our result is a factor $q_0$ smaller. The basic difference of our derivation from that of Ref.~\cite{swartz84}, is that we take into account that the length of a magnetic island flux tube is $Z_0 \simeq 2\pi q_0 R_0$, whereas in Ref.~\cite{swartz84} a flux tube with length $Z_0 \simeq 2 \pi R_0$ is considered, which is the major circumference of the tokamak. In other words, we point out that the magnetic island flux tube performs $q_0$ loops before closing on itself, and therefore all periodicity boundary conditions have to be imposed in its whole length, and not in the length of the tokamak circumference. This difference enters in particular the definition of the coordinate along the axis of the magnetic island flux tube $\zeta$, and approximated in Ref.~\cite{swartz84} by the toroidal coordinate $\zeta_T$.

In this work, we want to study the shear Alfv\'en wave propagation near the resonant flux surfaces where the energy is absorbed by continuum damping; therefore, we focus on the dynamics of modes that are characterized by radial singular structures. The linear equation for radially localized shear Alfv\'en modes in a compressible nonuniform tokamak plasma can be written in the form~\cite{chu92,turnbull93}:
\begin{equation}\label{eq:continuum-modes}
\frac{\omega^2}{\omega_A^2}\nabla_\perp^2\phi + Z_0^2 \nabla_\parallel \nabla_\perp^2 \nabla_\parallel \phi - \frac{\omega_{BAE-CAP}^2}{\omega_A^2} \nabla_\perp^2\phi= 0
\end{equation}
where $\omega_A = v_A/Z_0$. The frequency of the low-frequency SAW continuum accumulation point, delimiting the frequency gap of the Beta induced Alfv\'en Eigenmode (BAE), is denoted as $\omega_{BAE-CAP}$. We adopt, here, the value of the BAE-CAP given in reference~\cite{zonca96} and valid in the fluid limit, i.e. for frequencies much larger than the ion transit frequency:
\begin{equation}\label{eq:BAE}
\omega_{BAE-CAP} = \frac{1}{R_0} \sqrt{\frac{2 T_i}{m_i}\Big( \frac{7}{4} + \frac{T_e}{T_i}  \Big)}
\end{equation}
where $T_i$ and $T_e$ are the ion and electron temperatures and $m_i$ is the ion mass.
We focus on frequencies higher than $\omega_{BAE-CAP}$ and consistently neglect kinetic effects associated with wave-particle resonances~\cite{zonca96}. The operators $\nabla_{\parallel}$ and $\nabla_{\perp}$ are parallel and perpendicular gradients with respect to the equilibrium magnetic field.

In the following sections, the model equation for radially localized shear Alfv\'en modes, Eq.~\ref{eq:continuum-modes}, is written in the coordinate systems inside and outside the magnetic island and the numerical solution is provided and compared with analytical expressions.

\section{Solution inside the magnetic island}
\label{sec:inside}

\subsection{Eigenvalue problem}
Here, we write the equation for SAW continuum modes, Eq.~\ref{eq:continuum-modes}, in the coordinates inside the island, ($x,\theta,\zeta$), introduced above. We consider continuum modes with the same helicity of the magnetic island: $\partial /\partial \zeta = 0$. In this framework, the problem reduces to 2 dimensions. The parallel and perpendicular differential operators have the following form:
\begin{eqnarray}
Z_0 \nabla_\parallel & = & \sqrt{M} n_{isl} F \frac{\partial}{\partial \theta} \nonumber \\
\rho_{sx}^2 \nabla^2_\perp & = & a \frac{\partial^2}{\partial x^2} \nonumber
\end{eqnarray}
and the boundary conditions of the problem are periodicity conditions in $\theta$ on a $2\pi$ circle. By applying the differential operators in this explicit form, Eq.~\ref{eq:continuum-modes} can be written in the form of an eigenvalue problem:
\begin{equation}\label{eq:eigenvalue-inside}
\Big[ \frac{\Omega^2}{M n_{isl}^2} + \frac{F}{a} \frac{\partial}{\partial \theta} a F \frac{\partial}{\partial \theta}\Big] f = 0
\end{equation}
where $\Omega^2 = (\omega^2-\omega_{BAE-CAP}^2)/\omega_A^2$ is the eigenvalue, and $f = \partial^2 \phi / \partial^2 x $ is the eigenfunction.

\begin{figure}[t!]
\begin{center}
\includegraphics[width=0.55\textwidth]{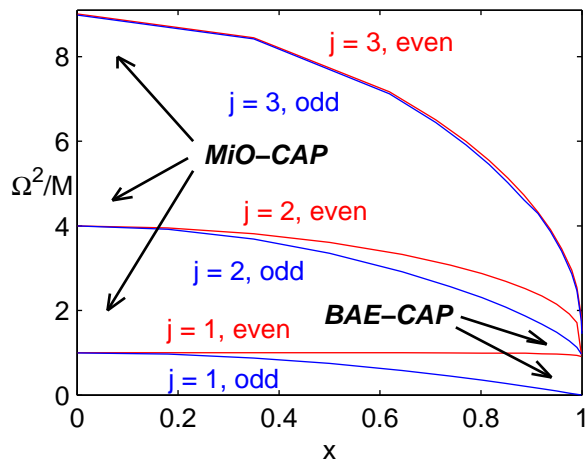}
\caption{Continuous spectrum $\Omega^2(x)$ for the small eccentricity case, $e \ll 1$, corresponding to $M\simeq 1$, plotted versus the radial position inside the island. Typical values of the equilibrium parameters have been chosen and $n_{isl}=1$. The O-point is at $x = 0$ and the separatrix at $x=1$. The branches of both even and odd eigenfunctions are shown. At the O-point, the continuum accumulation points (MiO-CAP) recover the value of the cylinder limit. At the separatrix, the first odd frequency branch tends to the original BAE-CAP ($\Omega^2 = 0$) and the others to the nonlinearly modified BAE-CAP ($\Omega^2 = M n_{isl}^2$).}
\label{fig:e0-n0-inside}
\end{center}
\end{figure}

It is useful to write Eq.~\ref{eq:eigenvalue-inside} in the form of a Schr\"odinger equation. The motivation is twofold. Firstly, we can use standard shooting method techniques to solve numerically the problem in the whole range inside the island: $0<x<1$. Secondly, we can solve analytically the problem in the proximity of the island O-point, approximating the potential, and compare the numerical solution with the analytical one near the O-point. The eigenvalue problem takes the form:
\begin{equation}\label{eq:schroedinger_in}
\frac{\partial^2}{\partial \theta^2} h + \frac{\Omega^2}{M n_{isl}^2} A_{in} h - V_{in} h = 0
\end{equation}
with $A_{in}= 1/F^2$, $V_{in} = (g_{in}'' - g_{in}'^2/(2g_{in}))/(2g_{in})$, $g_{in} = a F$, and $h = f\sqrt{g_{in}}$. Here the prime denotes the $\theta$-derivative, e.g. $g' = \partial g / \partial \theta$.

The solution of Eq.~\ref{eq:schroedinger_in}, can be found numerically with a shooting method code in the range $0<x<1$, for any value of the eccentricity $0<e<1$. The result is shown in Fig.~\ref{fig:e0-n0-inside} for $e\ll 1$, and in Fig.~\ref{fig:Msmall-summary} for $e = 0.99$. We choose typical values for the equilibrium parameters, $q_0 = 2$, $s=1$, $\varepsilon_0 = 0.1$, $n_{isl} = 1$. The periodic boundary condition in $\theta$ is satisfied by an infinite set of solutions, labeled by $j=1,2,...$. Three main features describe the continuous spectrum inside the island. 1) The continuous spectrum branches $\Omega^2_j$ have continuum accumulation points at the O-point, named here MiO-CAP. 2) Even and odd eigenfunction have different eigenvalues, due to the non-uniformity of the magnetic field intensity along the field line. 3) At the separatrix, the continuum frequencies converge to two different BAE-CAP. The first odd eigenfunction has frequencies converging to the original BAE-CAP, whose value is the same as in an equilibrium without islands, correspondent to $\Omega^2 = 0$. All other branches converge to a nonlinearly modified BAE-CAP, whose value is $\Omega = M n_{isl}^2$.

The solution of Eq.~\ref{eq:schroedinger_in} can also be found analytically near the O-point and near the separatrix, approximating the potential and obtaining the value of the continuum accumulation points (MiO-CAP and BAE-CAP). The analytical derivation of the MiO-CAP value is shown hereafter, and an estimation of the nonlinearly modified BAE-CAP is given in Section~\ref{subsec:separatrix}.

\subsection{Analytic solution near the O-point}
\label{sec:O-point}

The value of the continuum accumulation points at the O-point of the island can be calculated analytically from Eq.~(\ref{eq:schroedinger_in}), by approximating the functions $A_{in}$ and $V_{in}$ near the O-point. In this limit, Eq.~(\ref{eq:schroedinger_in}) becomes:
\begin{equation}\label{eq:schroedinger_in_O-point}
\frac{\partial^2}{\partial \theta^2} h + \frac{\Omega^2}{M n_{isl}^2} \, h - V_{in,0} h = 0
\end{equation}
where the potential is defined by:
\begin{equation}\label{eq:potential_O-point}
V_{in,0} (\theta) = e \, \Big( \frac{(1-e) \cos^2\theta}{(1 - e \cos^2\theta)^2} (2- e \cos^2\theta) \,- \, 1 \Big)
\end{equation}
Two limiting cases are considered here: $e \ll 1$, corresponding to $M\simeq 1$, and $e \simeq 1$, corresponding to $M \ll 1$. The former describes a  magnetic island where the flux surfaces have circular cross section near the O-point. The latter is representative of typical size magnetic islands in tokamak plasmas. In the former case, the function $a$ can be approximated by $a \simeq 1$ and the magnetic field intensity is independent of $\theta$: $
B_{ph}^\theta \simeq  \varepsilon_0 |s| B_0 \rho / (q_0 r_0 \gamma^2)$.
This means that, in our model, the $e\ll 1$ case has a cylindrical symmetry around the O-point. In this case, the potential $V_{in,0}$ vanishes and Eq.~\ref{eq:schroedinger_in_O-point} has the following eigenvalues:
\begin{equation}\label{eq:MiAE-CAP-cylinder-1}
 \Omega^2_{MiO-CAP}  =  M n_{isl}^2 j^2  = j^2/q_{in}^2(0)   \;\;\;\;\;\;\;\; ( \mathrm{case}\,\,\, e\ll 1)
\end{equation}
where $j$ is a natural number ($j=1,2...$), and $q_{in}(0)$ is the safety factor given in Eq.~\ref{eq:safety-factor}, calculated at $x=0$.

On the other hand, in the case of small island amplitude ($e\simeq 1$), the potential can be written as $V_{in,0}  \simeq V_0 + \delta V$, where $V_0  = -1$ and $ \delta V = V_1 \, H(\theta_{eff})$. Here $V_1 = 1/(1-e) = s^2 / (M n_{isl}^2 \gamma^2)$ and $H(\theta_{eff})$ is a function with unitary value inside the set $(|\theta| < \theta_{eff} ) \bigcup (|\pi - \theta| < \theta_{eff})$ and zero elsewhere, where $\theta_{eff} = (1/2) \, (\sqrt{M}/|s|)^{1/2} $. Since the potential $V_{in,0}$ is not constant in $\theta$, in this case the eigenvalue $\Omega^2$ has a different value for even and odd eigenfunctions. However, an approximated value of $\Omega^2$ can be calculated  by making some considerations on the potential shape. In fact, having $\delta V$ a support which is very localized in $\theta$ for small amplitude islands, where odd modes are small, we can neglect its contribution and consider the problem determined by $V_{in,0}=V_0 = -1$. This gives the estimate for odd mode eigenvalues:
\begin{equation}
\Omega^2_{MiO-CAP}  =  M n_{isl}^2 (j^2-1)     \;\;\;\;\;\;\;\; ( \mathrm{case}\,\,\, e \simeq 1)
\end{equation}

\section{Solution outside the magnetic island}
\label{sec:outside}

In this section, we face the problem of SAW continuum modes, described by  Eq.~\ref{eq:continuum-modes}, outside the separatrix of a magnetic island, in the coordinate system, ($x,u,\zeta$). The parallel and perpendicular differential operators have the following form:
\begin{eqnarray}
Z_0 \nabla_\parallel & = & 2 \sqrt{M} n_{isl} L \frac{\partial}{\partial u} \nonumber \\
\rho_{sx}^2 \nabla^2_\perp & = & \frac{P^2}{x^2} \frac{\partial^2}{\partial x^2} \nonumber
\end{eqnarray}
Only continuum modes with $m = q_0 n$ are considered, where $m$ and $n$ are respectively the poloidal and toroidal mode numbers, in the tokamak coordinates $\theta_T$ and $\zeta_T$. Those are the modes with the same helicity of the magnetic island: $\partial /\partial \zeta = 0$, and therefore, the problem reduces to 2 dimensions. By applying the differential operators in this explicit form, Eq.~\ref{eq:continuum-modes} can be written as in Sec.~\ref{sec:inside} as an eigenvalue problem:
\begin{equation}\label{eq:eigenvalue-outside}
\Big[ \frac{\Omega^2}{M n_{isl}^2} + \frac{4 L}{P^2} \frac{\partial}{\partial u} L P^2 \frac{\partial}{\partial u}\Big] f = 0
\end{equation}
and, as for Eq.~\ref{eq:eigenvalue-inside}, this can be put in the form of a Schr\"odinger equation. The Schr\"odinger equation can be solved numerically for $x>1$ with standard techniques and the numerical solution can be compared with the asymptotic limit, obtained analytically for $x\gg 1$. Moreover, analytical estimations of the frequency can be found near the separatrix. The eigenvalue problem takes the form:
\begin{equation}\label{eq:schroedinger_out}
\frac{\partial^2}{\partial u^2} h + \frac{\Omega^2}{M n_{isl}^2} A_{out} h - V_{out} h = 0
\end{equation}
with $A_{out}= 1/(4 L^2)$, $V_{out} = (g_{out}'' - g_{out}'^2/(2g_{out}))/(2g_{out})$, $g_{out} = 2 L P^2 $, and $h = f\sqrt{g_{out}}$.

\begin{figure}[t!]
\begin{center}
\includegraphics[width=0.55\textwidth]{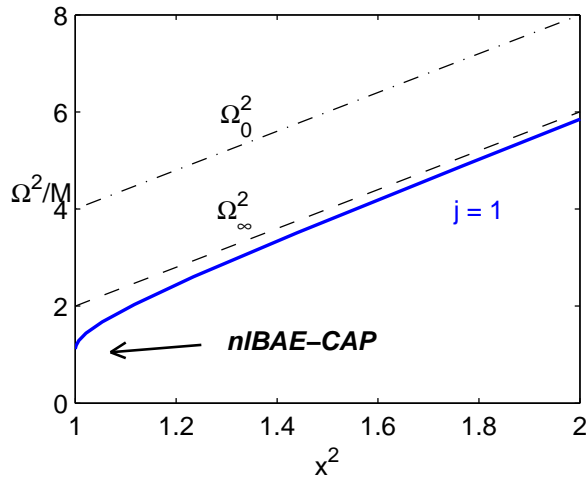}
\caption{Continuous spectrum $\Omega^2$, plotted versus $x^2$, for $M=10^{-2}$, corresponding to $e\simeq 0.99$. Typical equilibrium parameters have been chosen, and $n_{isl}=1$. Outside the magnetic island, all the frequency branches converge to a nonlinearly modified BAE-CAP at the separatrix ($x=1$), with value $\Omega^2_{nlBAE-CAP}= M n_{isl}^2$. On the other hand, for $x\gg 1$, the asymptotic limit $\Omega^2_\infty$ is approached. In our model, due to the constant-$\psi$ approximation, the solution without magnetic island $\Omega^2_0$ is shifted with respect to the asymptotic limit: $\Omega^2_0 = \Omega_\infty^2 + 2M$. Even and odd modes have a negligible difference in frequency: $\Delta \Omega^2 \sim O(M^2)$.}
\label{fig:Msmall-outside}
\end{center}
\end{figure}

\begin{figure}[t!]
\begin{center}
\includegraphics[width=0.55\textwidth]{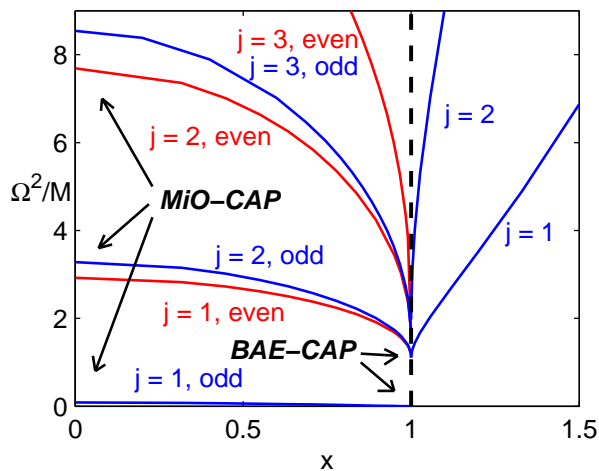}
\caption{Continuous spectrum $\Omega^2(x)$, for $M=10^{-2}$, corresponding to $e\simeq 0.99$. Typical equilibrium parameters have been chosen, and $n_{isl}=1$. The region inside the island is at $0<x<1$, and the region outside the island at $x>1$. The MiO-CAP are shown at the O-point ($x=0$), and the linear and nonlinear BAE-CAP at the separatrix ($x=1$).}
\label{fig:Msmall-summary}
\end{center}
\end{figure}

The periodic boundary condition in $u$ is satisfied by an infinite set of solutions, labeled with mode number $j$ (for $q_0$ integer, $j=n$, with $n$ the toroidal number in the tokamak $\zeta_T$ coordinate). The solution of Eq.~\ref{eq:schroedinger_out} is shown in Fig.~\ref{fig:Msmall-outside} and Fig.~\ref{fig:Msmall-summary}, for typical island size, $M = 10^{-2}$, and  typical values of the equilibrium parameters, $q_0 = 2$, $s=1$, $\varepsilon_0 = 0.1$, $n_{isl} = 1$. The asymptotic behavior at $x \gg 1$ is depicted in Fig.~\ref{fig:Msmall-outside} for the case $j=1$.   Three main features describe the continuous spectrum outside the island. 1) The continuous spectrum for $x\gg 1$ reaches the asymptotical limit $\Omega^2_{\infty}$, with a constant difference with the limit of the continuous spectrum in the limit of vanishing island, $\Omega^2_0$:
\begin{eqnarray}
 \Omega^2_{\infty} & = & n^2 <(q_T -q_0)^2>_u =  M n_{isl}^2 j^2 (4 x^2 -2)  \nonumber \\
 \Omega^2_0 & = &  n^2 \lim_{M\rightarrow 0} (q_T -q_0)^2 =M n_{isl}^2 j^2 (4 x^2)  \nonumber
\end{eqnarray}
This difference is due to the constant-$\psi$ approximation which is made modeling the island magnetic field. A more realistic asymptotic behavior can be obtained  by abandoning the constant-$\psi$ approximation and considering an improved model where the island field decays with growing $|q_T -q_0|$. 2) Even and odd eigenfunction have different eigenvalues, due to the non-uniformity of the magnetic field intensity along the field line, but the difference is negligible with respect to the absolute value. In the case depicted in Fig.~\ref{fig:Msmall-outside}, the difference between odd and even solutions is $\Delta \Omega^2 \sim O(M^2)$.
3) At the separatrix, all continuum frequencies converge to the nonlinearly modified BAE-CAP, whose value is proportional to the magnetic island half-width:
\begin{equation}\label{eq:BAE-CAP2}
\Omega_{nlBAE-CAP} = \sqrt{M} n_{isl}  = \frac{q_0 s n_{isl}}{2} \frac{W_{isl}}{r_0}
\end{equation}
This is consistent with the qualitative asymptotic behavior of the continuous spectrum near the separatrix, which can be calculated by knowing the value of the safety factor outside the magnetic island (see Eq.~\ref{eq:q-sx-approx}). Note that the nlBAE-CAP has the same frequency as the MiAE gap central frequency~\cite{Biancalani2010prl}.

The nonlinear modification of the continuous spectrum and in particular the upward shift in frequency of the BAE-CAP has important implications in the study of the dynamics of discrete AE in tokamak equilibria in the presence of a magnetic island, as discussed in Sec.~\ref{sec:conclusions}. The dynamics of discrete AE is not treated here, and will be discussed in a dedicated paper.  The analytical treatment of the continuous spectrum problem near the separatrix is presented in next section.

\subsection{Analytical treatment near the separatrix}
\label{subsec:separatrix}

Here, we focus on the region outside the magnetic island, in the proximity of the separatrix. We also consider the limit of $e\simeq 1$, which is the case of a typical size magnetic island in tokamak plasmas. Under these assumptions, the function $L$ and $P$ can be approximated as $L = P = \sqrt{1-\cos u}/\sqrt{2}$.
Consequently, the eigenvalue equation, Eq.~\ref{eq:schroedinger_out}, takes the form:
\begin{equation}
\frac{\partial^2}{\partial u^2} h - W_{\Omega^2} h = 0
\end{equation}
where the potential $W_{\Omega^2}$ is defined as:
\begin{displaymath}
 W_{\Omega^2} = - \frac{9 \cos^2 u - (12 + 8 \tilde\Omega^2) \cos u + (3 + 8 \tilde\Omega^2)}{16 (1- \cos u)^2}
\end{displaymath}
and $\tilde\Omega^2 = \Omega^2/(M n_{isl}^2)$.
This equation can be studied as a Schr\"odinger equation with zero energy. In this framework, we look for values of $\tilde\Omega^2$ such that the solution satisfies the boundary conditions of periodicity in $u$ on $(0, 2\pi)$. The potential $W_{\Omega^2}$ has positive second derivative for $\tilde\Omega^2 < 3/4 $, and negative second derivative for  $\tilde\Omega^2 > 3/4 $. For $\tilde\Omega^2 = 3/4 $, the potential has a constant value: $W_{\Omega^2} = -(3/4)^2$. For the given boundary conditions, this means that solutions of the problem exist for $\tilde\Omega^2 > 3/4 $ only. On the other hand, when $\tilde\Omega^2 = 4/3$ the potential has maximum value $W_{\Omega^2} = -1$ at $u=\pi$, and therefore does not admit a solution with mode $j=1$. With these considerations, we can estimate that $3/4 < \tilde\Omega^2_{nlBAE-CAP} < 4/3$ for the first mode number, $j=1$. Numerically, we find that $\tilde\Omega^2 \simeq 1$ for $j=1$ and that the difference in frequency for other mode numbers and different parities is negligible.

\section{Conclusions and discussion}
\label{sec:conclusions}

Understanding the structure of the SAW continuous spectrum is one crucial step in studying the stability properties of a tokamak plasma. In fact, one of the main damping mechanisms of SAW instabilities in non-uniform plasmas is continuum damping, which occurs near the magnetic flux surfaces where the frequency of the instability matches the SAW continuum frequency. In toroidal devices, a frequency gap in the continuous spectrum represents a range of frequency where discrete Alfv\'en eigenmodes (AE) can grow unstable, practically unaffected by continuum damping.

We have studied the radial structure of the SAW continuous spectrum in the presence of a magnetic island. Since magnetic islands are very common in fusion plasma experiments, for example in connection with tearing instabilities and sawtooth activity, the solution of this problem has important implications for understanding the stability regimes of AE in a fusion reactor. We have adopted a linear MHD model for radially localized SAW, where tokamak curvature is neglected except for the creation of the beta induced Alfv\'en eigenmode (BAE) frequency gap, in the low-frequency range of the SAW continuous spectrum. We also assumed modes with the same helicity of the magnetic island. In this framework, the model reduces to 2 dimensions. The case of local singular structures with helicity different from that of the magnetic island is considered in a different paper~\cite{Biancalani2010prl}. Due to the time scale separation between island and the SAW dynamics, the island has been treated as a static helical distortion of the equilibrium.

We have shown that the SAW continuous spectrum of a finite-$\beta$ tokamak plasma is modified by the presence of a magnetic island. In particular, the BAE continuum accumulation point (BAE-CAP) is shifted in space from the island rational surface, to the separatrix flux surface position.
Moreover, we have shown that the nonuniformity of the magnetic field intensity along the field lines generates a frequency splitting between modes with even and odd parity. This splitting frequency has a finite value inside the island and is negligible outside for typical island sizes.
Inside the magnetic island, we have found that the continuum frequency branches converge at the O-point to several magnetic island induced CAP (MiO-CAP). At the separatrix, the frequency of the BAE-CAP is split in two values. The continuum branch corresponding to the first odd eigenfunction inside the island converges to the original BAE-CAP with same frequency as without magnetic island; and the other branches converge to a nonlinearly modified BAE-CAP, with a higher frequency given by Eq.~\ref{eq:BAE-CAP2}.
Outside the magnetic island, all continuum branches at the separatrix converge to the nonlinearly modified BAE-CAP. Far from the separatrix and outside the magnetic island, the continuum frequency converges asymptotically to the value calculated in an equilibrium with no island.

Our results have two main implications in the study of the dynamics of Alfv\'en eigenmodes (AE). Inside the island, new magnetic island induced AE (MiAE) could exist as bound states, essentially free of continuum damping, provided that plasma equilibrium effects and free energy sources can drive and bind them locally (see also reference~\cite{Biancalani2010prl}). Outside the island,  BAE are expected to have an eigenfunction radially peaked in the proximity of the magnetic island separatrix, namely at the BAE-CAP, other than at the rational surface. The BAE frequency is expected to have higher values when a magnetic island is present, consistently with the BAE-CAP frequency modification described by Eq.~\ref{eq:BAE-CAP2}. Moreover, while inside the magnetic island the pressure and temperature profiles are flattened, outside the magnetic island, in the proximity of the separatrix, the equilibrium pressure gradients are highly increased. This may provide a free-energy source for BAE, which are expected to grow unstable in this configuration, even in absence of fast particle drive. BAE are generally damped by thermal ion Landau damping; therefore, we may expect to detect BAE when the value of the pressure gradient is over a certain threshold, namely when the magnetic island exceeds a certain size. The dispersion relation and the drive and damping mechanisms of BAE in the presence of a magnetic island will be treated in a different paper.

Experimental observations of Alfv\'enic modes in the presence of a magnetic island have been recently reported in several tokamaks devices
(see for instance the observations in FTU~\cite{buratti05nufu,BiancalaniIAEA08} and in HL-2A~\cite{Wei2010}), even in purely Ohmic heated plasmas.
The fluctuations frequency range confirms that they are BAE nonlinearly interacting with the magnetic island, characterized by a frequency shift dependence on the island size consistent with that of the nonlinearly modified BAE-CAP. Moreover, BAE were observed only above a certain threshold in the magnetic island amplitude.

The dependence on the magnetic island size of the continuous spectrum and in particular the MiO-CAP and the BAE-CAP suggests the possibility of using the MiAE and BAE frequency scalings as novel magnetic island diagnostics in a fashion similar to other commonly used Alfv\'en spectroscopy techniques~\cite{Breizman05,Zonca09,Pinches04}. The radial MiAE localization at the center of the island makes them more difficult to detect by external measurements than BAE and makes the use of internal fluctuation diagnostics, such as electron cyclotron emission (ECE) and soft X-Rays, necessary. On the other hand, BAE are easier to detect with external diagnostic techniques, and therefore, the possibility of using BAE frequency scalings as a magnetic island diagnostic can be considered more feasible.

\section*{Acknowledgments}
This work was supported by the Euratom Communities under the contract of Association between EURATOM/ENEA and in part by PRIN 2006 and CREATE, as well as by DOE grants DE-FG02-04ER54736 and DE-FC02-04ER54796, and by the NSF grant ATM-0335279.
Part of this work was done while one of the authors, A.B., was at the University of Pisa, in collaboration with University of California, Irvine, and ENEA Frascati Research Center, which are gratefully acknowledged for the hospitality. One of the author, A.B. would also like to thank Andreas Bierwage for the precious suggestions in writing the shooting method code. Useful discussions with C. Di Troia, I. Chavdarovski, X. Wang and P. Porazik are also acknowledged.

\appendix

\section{Coordinate metric}
\label{appendix:metric}

Here, we provide the metric for the coordinates used in this work outside the magnetic island $(\rho,u,\zeta)$ and inside the magnetic island $(\rho,\theta,\zeta)$. This is necessary to calculate the differential operators in the equation for the dynamics of continuum modes.

The domains of the coordinates $(\rho,u,\zeta)$, describing the region outside the island, are $\rho_{sx}= W_{isl} < \rho < \infty$, $0 < u < 2\pi$ and $0 < \zeta < 2\pi$, where $\zeta$ is the coordinate of translational symmetry for both equilibrium and perturbations, and periodicity in $u$ is assumed for the perturbations. The gradients of these coordinates are:
\begin{eqnarray}
\bm{\nabla} \rho & = & L/x  \, \hat{\bm{q}}_T - \sqrt{1-e}\sin u /(2 x)  \, \hat{\bm{u}} = (P/x) \, \hat{\bm{\rho}}\nonumber \\
\bm{\nabla} u & = & \hat{\bm{u}} /\rho_0 \nonumber \\
\bm{\nabla} \zeta & = & \hat{\bm{\zeta}}/ Z_0 \nonumber
\end{eqnarray}
where $\rho_0 = r_0/(q_0 n_{isl} \gamma)$, $Z_0 = \gamma q_0 R_0$ and we use the notation $\hat{\bm{V}} = \bm{V}/V$ for a versor (unit vector). Here $P = \sqrt{L^2 + (1-e)(\sin^2 u)/4}$, where $L = \sqrt{x^2 - (\cos u + 1)/2}$ and $x = \rho/\rho_{sx}$.

A covariant basis ($\bm{g}_\rho ,\bm{g}_u, \bm{g}_\zeta$) can also be defined as orthogonal to the contra-variant basis ($\bm{g}^\rho = \bm\nabla \rho ,\bm{g}^u = \bm\nabla u , \bm{g}^\zeta = \bm\nabla \zeta$), namely satisfying: $\bm{g}^i \cdot \bm{g}_j = \delta^i_j$. With this notation, any vector can be written as $\bm{V}= \sum_i V^i \bm{g}_i = \sum_i V^i_{ph} \hat{\bm{g}}_i$, where we call $V^i_{ph}$ the \emph{physical} components of the vector, $V^i_{ph} = V^i |\bm{g_i}|$.
The Jacobian for the coordinates outside the island, defined as the determinant of the metric tensor $g_{ij}= \bm{g}_i \cdot \bm{g}_j$, is $g_{out} = \rho_0^2 Z_0^2 x^2 / L^2$. With these definitions, the laplacian of a function $f$ for ``radially'' localized modes, can be written in the coordinates $(\rho,u,\zeta)$  as:
\begin{equation}
\nabla^2_{out} f \simeq \frac{1}{\sqrt{g_{out}}} \frac{\partial}{\partial \rho} \Big( \sqrt{g_{out}} \, g^{\rho \rho} \frac{\partial f}{\partial \rho} \Big) \simeq \frac{1}{\rho_{sx}^2} \frac{P^2}{x^2} \frac{\partial^2 f}{\partial x^2}
\end{equation}
where we have used $g^{\rho \rho} =  \bm{g}^\rho \cdot \bm{g}^\rho = P^2 / x^2$.

The domains of the coordinates $(\rho,\theta)$, for the region inside the island, are $0 < \rho < \rho_{sx}= W_{isl}$, $0 < \theta < 2\pi$ and $0 < \zeta < 2\pi$, where $\zeta$ is the coordinate of translational symmetry as above, and periodicity in $\theta$ and $\zeta$ is assumed for the perturbations. The gradients of the coordinates $\rho$ and $\theta$ are:
\begin{eqnarray}
\bm{\nabla} \rho & = & \sin\theta \, \hat{\bm{q}}_T - \cos\theta \sqrt{1-e}  F  \, \hat{\bm{u}} = \sqrt{a} \, \hat{\bm{\rho}}\nonumber \\
\bm{\nabla} \theta & = & ( \cos\theta \, \hat{\bm{q}}_T + \sin\theta \sqrt{1-e}  F  \, \hat{\bm{u}} ) /\rho  = \sqrt{b} \, \hat{\bm{\theta}} /\rho \nonumber
\end{eqnarray}
Moreover we have the relation $\bm{\nabla} \rho \cdot \bm{\nabla} \theta = c/\rho $. Here $F = \sqrt{1 - x^2 \cos^2\theta}$, $ a = \sin^2\theta + \cos^2\theta (1-e)  F^2$, and $c=\cos\theta \sin\theta \,(1 - (1-e) F^2)$. The Jacobian for the coordinates inside the island, is $g_{in} = \rho_{sx}^2 Z_0^2 x^2 / (F^2 (1-e))$. Using these definitions, and the fact that $g^{\rho \rho} =  \bm{g}^\rho \cdot \bm{g}^\rho = a$, we write the laplacian for ``radially'' localized modes in the coordinates $(\rho,\theta,\zeta)$  as:
\begin{equation}
\nabla_{in}^2 f  \simeq \frac{a}{\rho_{sx}^2} \frac{\partial^2 f}{\partial x^2}
\end{equation}


\section*{References}

\begin{thebibliography}{99}

\bibitem{Rosenbluth75} Rosenbluth M N and Rutherford P H 1975 {\it Phys. Rev. Lett.} {\bf 34} 1428
\bibitem{Mikhailovskii75} Mikhailovskii A B 1975 {\it Zh. Eksp, Teor. Fiz.} {\bf 68} 1772
\bibitem{Chen89} Chen L 1989, in ``Theory of Fusion Plasmas'', Eds J. Vaclavick (Bologna: SIF) p.327
\bibitem{Fu89} Fu G Y and Van Dam J W 1989 {\it Phys. Fluids} {\bf B1} 1949
\bibitem{Zonca06b} Zonca F {\sl et al.} 2006 {\it Plasma Phys. Control. Fusion} {\bf 48} B15
\bibitem{chen07} Chen L and Zonca F 2007 {\it Nucl. Fusion} {\bf 47} S727
\bibitem{Grad69} Grad H 1969 {\it Phys. Today} {\bf 22} (12) 34
\bibitem{hasegawa74} Hasegawa A and Chen L 1974 {\it Phys. Rev. Lett.} \textbf{32} 454
\bibitem{chen74} Chen L and Hasegawa A 1974 {\it Phys. Fluids} {\bf 17} 1399
\bibitem{kieras82} Kieras C E and Tataronis J A 1982 {\it J. Plasma Physics} \textbf{28} 395


\bibitem{cheng85} Cheng C Z, Chen L and Chance M S 1985 {\it Ann Phys} \textbf{161} 21
\bibitem{Zonca92} Zonca F and Chen L 1992 {\it Phys. Rev. Lett.} {\bf 68} 592
\bibitem{Chen95} Chen L and Zonca F 1995 {\it Physica Scripta} {\bf T60} 81
\bibitem{Zonca06a} Zonca F and Chen L 2006 {\it Plasma Phys. Control. Fusion} {\bf 48} 537
\bibitem{chen94} Chen L 1994 {\it Phys. Plasmas} \textbf{1} 1519
\bibitem{Alfven50} Alfv\'en H 1950 ``Cosmical Electrodynamics'', Clarendon Press, Oxford
\bibitem{Elsasser56} Elsasser W M 1956 {\it Rev Mod Phys} \textbf{28} 135
\bibitem{Hasegawa89} Hasegawa A and Sato T 1989 ``Space Plasma Physics (Stationary Processes)'' Vol. 1 (New York; Springer)
\bibitem{furth63} Furth H P, Killeen J and Rosenbluth M N 1963 {\it Phys. Fluids} \textbf{6} 459
\bibitem{BiancalaniEPS08} Biancalani A, Chen L, Pegoraro F and Zonca F 2008 {\it Proc. 35th EPS Plasma Physics Conf.}, 9-13 June 2008, Hersonissos, Crete, Greece, Paper P1-051 http://epsppd.epfl.ch



\bibitem{Tuccillo09} Tuccillo A A {\sl et al.} 2009 {\it Nucl. Fusion} {\bf 49} 104013
\bibitem{chu92} Chu M S, Greene J M, Lao L L, Turnbull A D and Chance M S 1992 {\it Phys. Fluids B} {\bf 4} 3713
\bibitem{turnbull93} Turnbull A D {\it et al.} 1993 {\it Phys. Fluids B} {\bf 5} 2546
\bibitem{zonca96} Zonca F, Chen L and Santoro R A 1996 {\it Plasma Phys. Control. Fusion} {\bf 38} 2011
\bibitem{Biancalani2010prl} Biancalani A, Chen L, Pegoraro F and Zonca F 2010 {\it Phys. Rev. Lett.} \textbf{105} 095002
\bibitem{swartz84} Swartz K and Hazeltine R D 1984 {\it Phys. Fluids} \textbf{27} 2043
\bibitem{buratti05nufu} Buratti P {\it et al.} 2005 {\it Nuclear Fusion} {\bf 45} 1446
\bibitem{BiancalaniIAEA08} Biancalani A {\it et al.}, {\it Proc. 22nd IAEA Fusion Energy Conf.}, 13-18 October 2008, Geneva, Switzerland, Paper Th/p3-5, http://www-pub.iaea.org/MTCD/Meetings/fec2008pp.asp
\bibitem{annibaldi07} Annibaldi S V, Zonca F and Buratti P 2007 {\it Plasma Phys. Control. Fusion} {\bf 49} 475
\bibitem{Wei2010} Wei C {\it et al.} 2010 {\it Journal of the Phys. Soc. Japan} {\bf 79} 044501



\bibitem{Rutherford73} Rutherford P H 1973 {\it Phys. Fluids} {\bf 16} 1903--1908
\bibitem{Breizman05} Breizman B N, Pekker M S and Sharapov S E  2005 {\it Phys. Plasmas} {\bf 12} 112506
\bibitem{Zonca09} Zonca F {\it et al.} 2009 {\it Nucl. Fusion} {\bf 49} 085009
\bibitem{Pinches04} Pinches S D {\it et al.} 2004 {\it Plasma Phys. Control. Fusion} {\bf 46} B187


\end{thebibliography}
\end{document}